\documentclass[a4paper]{jpconf}
\usepackage{graphicx}
\usepackage{url}
\usepackage{amsmath}
\usepackage{array}

\usepackage{color}

\newcommand{\ND}{N_{\rm D}}
\newcommand{\NF}{N_{\rm F}}
\newcommand{\VD}{V_{\rm D}}

\newcommand{\rhoG}{\rho_{0}}
\newcommand{\rhoL}{\rho_{\rm L}}
\newcommand{\rhoC}{\rho_{\rm c}}
\newcommand{\tauW}{\tau_{\rm W}}
\newcommand{\DeltaC}{\Delta_{\rm c}}
\newcommand{\lambdaC}{\lambda_{\rm c}}

\newcommand{\kB}{k_{\rm B}}

\newcommand{\Ising}{^{\rm Is}}
\newcommand{\TI}{T\Ising}
\newcommand{\tauWI}{\tau_{\rm W}\Ising}
\newcommand{\mykappa}{\hat{\kappa}}
\newcommand{\betaI}{\beta\Ising}

\begin{document}
\title{Numerical test of finite-size scaling predictions for the droplet
condensation-evaporation transition}

\author{Andreas Nu{\ss}baumer$^1$, Johannes Zierenberg$^2$,  Elmar Bittner$^3$ and Wolfhard Janke$^2$}
\address{$^1$ Institut f\"ur Physik, Johannes Gutenberg Universit\"at
              Mainz, Staudinger Weg 7, D-55128 Mainz, Germany}
\address{$^2$ Institut f\"ur Theoretische Physik, Universit\"at Leipzig,
              Postfach 100\,920, D-04009 Leipzig, Germany}
\address{$^3$ Institut f\"ur Theoretische Physik, Universit\"at Heidelberg,
              Philosophenweg 16, D-69120 Heidelberg, Germany}
\ead{wolfhard.janke@itp.uni-leipzig.de}

\begin{abstract}
  We numerically study the finite-size droplet condensation-evaporation
  transition in two dimensions. We consider and compare two orthogonal
  approaches, namely at fixed temperature and at fixed density, making use of
  parallel multicanonical simulations. The equivalence between Ising model and
  lattice gas allows us to compare to analytical predictions. We recover the
  known background density (at fixed temperature) and transition
  temperature (at fixed density) in the thermodynamic limit and compare our
  finite-size deviations to the predicted leading-order finite-size corrections.
\end{abstract}

%%%%%%%%%%%%%%%%%%%%%%%%%%%%%%%%%%%%%%%%%%%%%%%%%%%%%%%%%%%%%%%%%%%%%%%%%%%%%%%%%%%%%%%%%%%%
%%%%%%%%%%%%%%%%%%%%%%%%%%%%%%%%%%%%%%%%%%%%%%%%%%%%%%%%%%%%%%%%%%%%%%%%%%%%%%%%%%%%%%%%%%%%
\section{Introduction}
Droplet formation is an essential process in nature with a variety of analogues
in biological systems and material science. Of course, this generally involves
non-equilibrium processes, e.g., the formation of nucleation prerequisites from
local fluctuations while the surrounding gas acts as a density bath. Here, we
consider instead a canonical setup in a finite system of size $V$ with fixed
temperature $T$ and particle number $N$. If the gas is supersaturated, a
variation of $N$ or $T$ results in the formation of equilibrium
droplets~\cite{binder1980, neuhaus2003, biskup2002, biskup2003, binder2003}. In
simple terms, the particle excess is subdivided to form a single macroscopic
droplet in equilibrium with the surrounding vapor plus remaining excess.
The resulting theory has been supported by numerous computational studies at
fixed temperature, including the two-dimensional lattice
gas~\cite{nussbaumer2006, nussbaumer2008, zierenberg2014jpcs} and
three-dimensional Lennard-Jones gas~\cite{macdowell2004, macdowell2006,
schrader2009}. The orthogonal approach at fixed density has received less
attention~\cite{martinos2007}, but recently enabled us to come closer to the
asymptotic scaling regime for two- and three-dimensional lattice gas and
three-dimensional Lennard-Jones gas~\cite{zierenberg2015}. In the following, we
will compare the leading-order scaling corrections for a two-dimensional lattice
gas at fixed temperature and fixed density from analytical predictions with
numerical results.

%%%%%%%%%%%%%%%%%%%%%%%%%%%%%%%%%%%%%%%%%%%%%%%%%%%%%%%%%%%%%%%%%%%%%%%%%%%%%%%%%%%%%%%%%%%%
%%%%%%%%%%%%%%%%%%%%%%%%%%%%%%%%%%%%%%%%%%%%%%%%%%%%%%%%%%%%%%%%%%%%%%%%%%%%%%%%%%%%%%%%%%%%
\section{Model}
\label{secModel}
We consider a lattice gas in $d=2$ dimensions. Excluded volume is modeled as
lattice sites being either occupied by exactly one particle or empty,
$n_i=\{0,1\}$. Short-range interaction is included by nearest-neighbor
interaction ($\langle i,j\rangle$). The Hamiltonian is then
\begin{equation}
  \mathcal{H}=-\sum_{\langle i.j\rangle} n_i n_j,
\end{equation}
which is equivalent to an Ising model at fixed magnetization
with $\TI = 4T$ and coupling constant $J=1$ \cite{lee1952}.
This originates in the identification of the spin state
$s_i=2n_i-1$, which allows one to rewrite the Ising Hamiltonian
\begin{equation}
  \mathcal{H}^{\rm Is}
  = -\sum_{\langle i,j\rangle} s_i s_j
  =  4\mathcal{H} - d(V-4N),
\end{equation}
considering that, on a simple hypercubic lattice, the sum over nearest neighbors
yields $d$ contributions and $N=\sum n_i$. The equivalence is then established
by equating the Boltzmann factors $\exp(-\beta\mathcal{H})$, where $\beta=(\kB
T)^{-1}$. Evaluating $\betaI\mathcal{H}^{\rm Is} = \beta\mathcal{H}$, where the
constant shift is neglected due to physical unimportance, results in a
simple rescaling of the temperature. When making use of the Ising equivalence
all energy-related observables have to be rescaled correspondingly. The
canonical Ising model is then equivalent to a grand-canonical lattice gas.

% relevant constants
Exploiting the equivalence to the 2D Ising model, the critical point is located
at $\beta_c=4\betaI_c = 2\ln\left(1+\sqrt{2}\right) \approx 1.763$ or $\kB
T_c\approx 0.567$. For temperatures below this point, a spontaneous
magnetization $m_0$ is observed and described by the Onsager-Yang
solution~\cite{onsager1949,yang1952}:
\begin{equation}
  m_0(\betaI) = \left[1- \sinh^{-4}(2\betaI)\right]^{1/8}.
  \label{eqM}
\end{equation}
This is directly related to the (grand-canonical) equilibrium background
density $\rho_0 = (1-m_0)/2$.
The magnetic susceptibility is connected with the isothermal compressibility
$\chi=\beta\kappa=\mykappa$~\cite{zierenberg2014jpcs} and may be evaluated from
sufficiently long series expansions (see, e.g., Ref.~\cite{boukraa2008jphysA}
and references therein), where
\begin{equation}
  \chi(\betaI) = \betaI\sum_{i=0}^{n} c_i u^{2i} \quad \mathrm{with}\quad
  u=\frac{1}{2\sinh(2\betaI)},
  \label{eqChi}
\end{equation}
and
$c=\{0,0,4,16,104,416,2224,8896,43840,175296,825648,3300480,15101920,...\}$,%
\footnote{
The coefficients were obtained from
\url{http://www.ms.unimelb.edu.au/~iwan/ising/Ising_ser.html}~\cite{boukraa2008jphysA}.
} here considered up to the $300^{\rm th}$ term.
The equilibrium shape of a 2D Ising droplet is described by the Wulff plot (or
shape), given by~\cite{leung1990}
\begin{equation}
  W = \frac{4}{(\betaI)^2}
  \int_0^{\betaI\sigma_0}\mathrm{d} x~\cosh^{-1}\left[\frac{\cosh^2(2\betaI)}{\sinh(2\betaI)}-\cosh(x)\right],
  \label{eqW}
\end{equation}
where $\sigma_0=2+\ln[\tanh(\betaI)]/\betaI$ and $\cosh^{-1}$ is referring to
the inverse hyperbolic cosine. This will be relevant for the surface free energy
of a (Wulff shaped) droplet of unit volume $\tauWI=2\sqrt{W}$. Being
energy-related, the interface tension gets converted as
$\tauW=\tauWI/4$.

%%%%%%%%%%%%%%%%%%%%%%%%%%%%%%%%%%%%%%%%%%%%%%%%%%%%%%%%%%%%%%%%%%%%%%%%%%%%%%%%%%%%%%%%%%%%
%%%%%%%%%%%%%%%%%%%%%%%%%%%%%%%%%%%%%%%%%%%%%%%%%%%%%%%%%%%%%%%%%%%%%%%%%%%%%%%%%%%%%%%%%%%%
\section{Method}
In order to obtain numerical data at fixed temperature and at fixed density, we
employed two different kinds of simulations. In both cases the underlying
algorithm is the multicanonical method~\cite{berg1991, berg1992, janke1992,
janke1998}. The condensation transition is a first-order phase transition for
which this method is well suited because it potentially allows one to overcome
barriers in the free energy. The principle idea is to replace the Boltzmann
weight $\exp(-\beta E)$ by an a priori unknown weight function $W(E)$, which is
iteratively adapted in order to yield a flat histogram over a desired energy
range. That way, energy states which are suppressed at a first-order phase
transition are artificially enhanced and the simulation may transit back and
forth between the involved coexisting phases. In the end, the simulation data
has to be reweighted to estimates of expectation values at any temperature for
which the probability distribution is covered by the flat histogram. The
``optimal'' weight function is reached if a simulation run produces a flat
histogram, i.e., for a fixed $c$ it holds $h(E_{\rm min})/h(E_{\rm max}) > c$.
% pmuca
In part, we further make use of a parallel
implementation~\cite{zierenberg2013cpc, zierenberg2014pp}, which exploits the
fact that in each iteration the histogram is an estimate of the probability
distribution belonging to the current weight function. This allows one to
distribute the sampling to independent Markov chains and to obtain a joint
estimate as a simple sum of individual histograms. The procedure scales very
well for the problem at hand~\cite{zierenberg2014jpcs}.

% technical details
The involved Monte Carlo updates include particle displacements to free nearest
neighbors and random jumps to free sites. This corresponds to local and global
Kawasaki updates. Errors are obtained using the Jackknife
method~\cite{efron1982} and standard error propagation. For further details of
the employed methods we refer to Refs.~\cite{nussbaumer2006, nussbaumer2008,
zierenberg2014jpcs, zierenberg2015}.

%%%%%%%%%%%%%%%%%%%%%%%%%%%%%%%%%%%%%%%%%%%%%%%%%%%%%%%%%%%%%%%%%%%%%%%%%%%%%%%%%%%%%%%%%%%%
%%%%%%%%%%%%%%%%%%%%%%%%%%%%%%%%%%%%%%%%%%%%%%%%%%%%%%%%%%%%%%%%%%%%%%%%%%%%%%%%%%%%%%%%%%%%
\section{Theory of leading-order correction}

A natural approach to particle condensation/evaporation is the consideration of
a grand-canonical ensemble at fixed temperature, where the system relaxes to an
equilibrium background contribution $N_0=\rho_0V$. Below the corresponding
critical temperature the system is dominated either by particles (fluid branch)
or by void space (gas branch) and $\rho_0\neq1/2$. Let us now consider the
dilute gas branch well below the critical temperature and fix $N>N_0$. This
results in the canonical ensemble of a supersaturated gas with particle excess
$\delta N=N-N_0$. Initially, this excess goes into the gas phase while for
sufficiently large excess droplet formation occurs.

\begin{figure}
  \centering
  \includegraphics[width=0.30\textwidth]{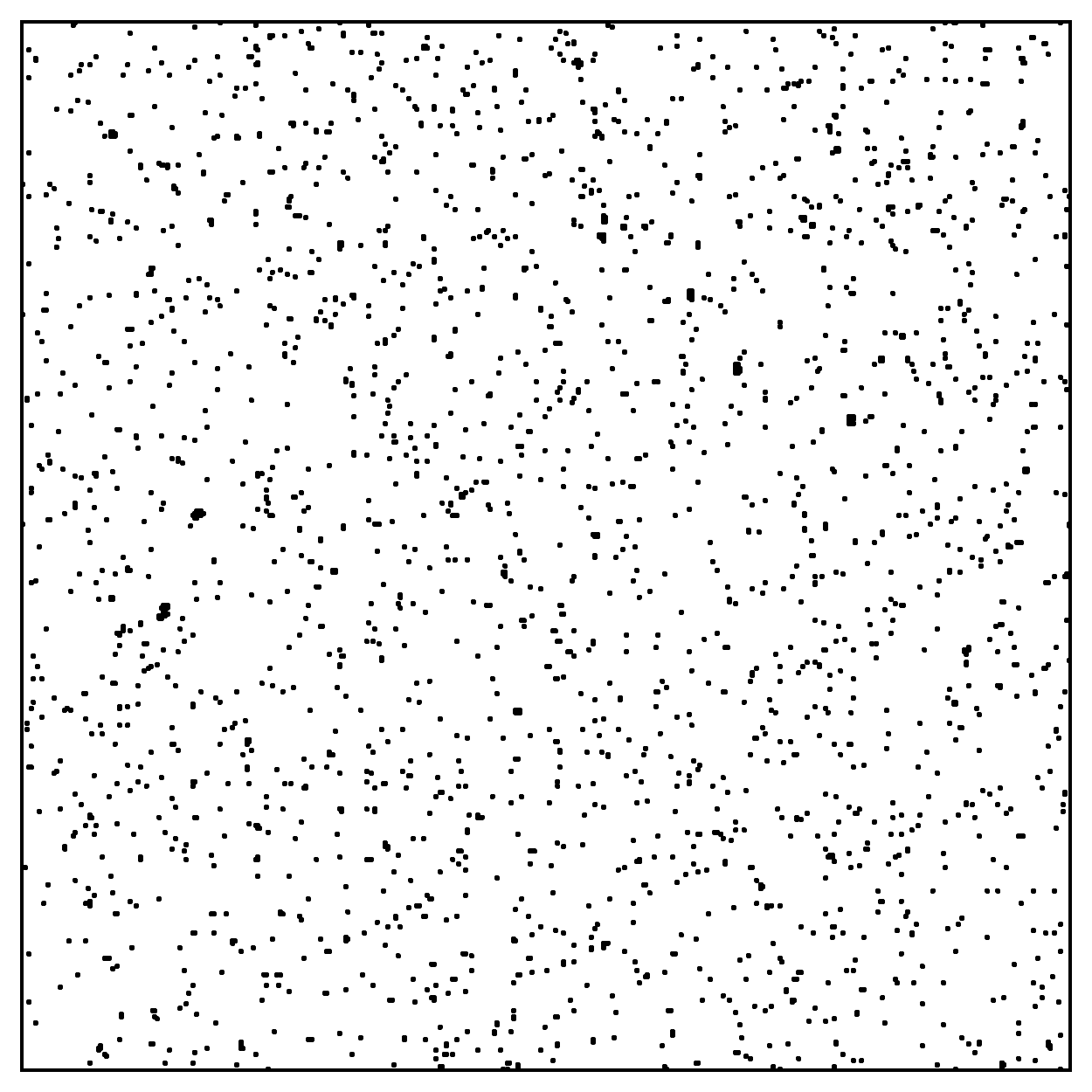}
  \hspace{6em}
  \includegraphics[width=0.30\textwidth]{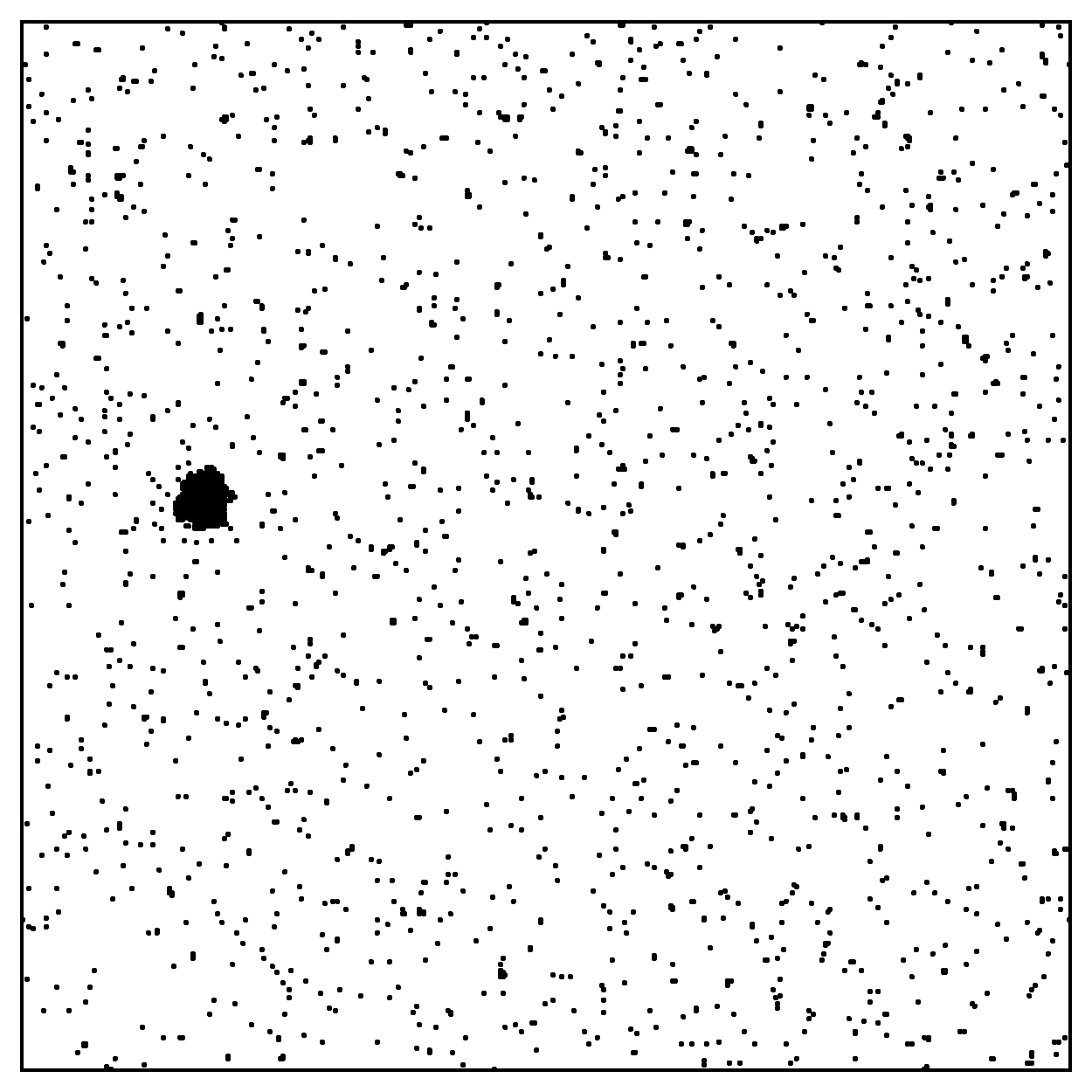}
  \caption{%
    Snapshots of a two-dimensional lattice gas at fixed density above (left) and
    below (right) the condensation-evaporation temperature, showing a homogeneous
    gas phase and a droplet in equilibrium with surrounding vapor, respectively.
    Both plots show $N=2500$ particles on a lattice of linear size $L=500$, i.e.,
    $\rho=0.01$.
    \label{figCondensationPhases}
  }
\end{figure}
In equilibrium droplet formation, the probability for intermediate-sized
droplets was shown to vanish~\cite{biskup2002} and the scenario reduces to a
homogeneous gas phase and an inhomogeneous phase of a droplet in equilibrium
with surrounding vapor. This may be considered as the interplay of entropy
maximization by fluctuations in the gas phase and energy minimization by forming
a droplet~\cite{biskup2002,biskup2003}, see also
Fig.~\ref{figCondensationPhases}. At fixed temperature it is possible to
consider (fixed) thermal fluctuations and relate them to infinite-size
temperature-dependent quantities, like the isothermal compressibility $\mykappa$
and the normalized surface free energy $\tauW$, see Sec.~\ref{secModel}.
The free energy may then be approximated by a contribution $F_{\rm fluc}$ from
the fluctuation of \mbox{particle excess $\delta N$} and a contribution $F_{\rm
drop}$ from the single macroscopic droplet of size $\VD$:
\begin{equation}
  F_{\rm fluc} = \frac{(\delta N)^2}{2\mykappa V}
  \qquad\text{and}\qquad
  F_{\rm drop} = \tauW(\VD)^\frac{d-1}{d}.
\end{equation}
These contributions are idealized with possible sources of corrections in both
the Gaussian approximation and the droplet shape for finite systems.

In the two-phase scenario, the particle excess may be decomposed into the excess
inside the droplet $\delta \ND$ and the excess in the fluctuating phase
$\delta\NF$, i.e.,
%
%\begin{equation}
  $\delta N = N-N_0 = \delta\ND + \delta\NF$.
%\end{equation}
%
Linking the droplet size to the particle excess inside the droplet, one expects
\mbox{$\delta\ND=(\rhoL -\rhoG)\VD$}, where $\rhoL$ and $\rhoG$ are the
background liquid and gas density, respectively. Then, one may introduce a
scalar fraction
\begin{equation}
  \lambda = \delta\ND/\delta N,
  \label{eqLambda}
\end{equation}
such that $\delta\ND=\lambda \delta N$ and $\delta\NF = (1-\lambda)\delta N$.
The total free energy $F=F_{\rm drop}+F_{\rm fluc}$ becomes
\begin{equation}
  F = \tauW \left( \frac{\lambda\delta N}{\rhoL-\rhoG}\right)^{\frac{d-1}{d}} + \frac{(1-\lambda)^2(\delta N)^2}{2\mykappa V}
    = \tau_W \left(\frac{\delta N}{\rhoL-\rhoG}\right)^{\frac{d-1}{d}}
       \left(\lambda^{\frac{d-1}{d}} + \Delta (1-\lambda)^2\right),
  \label{eqCondensationFixTFreeEnergy}
\end{equation}
with a dimensionless ``density'' parameter
\begin{equation}
  \Delta
  = \frac{(\rhoL-\rhoG)^{\frac{d-1}{d}}}{2\mykappa \tau_W} \frac{(\delta
  N)^{\frac{d+1}{d}}}{V}
  = \frac{(\rhoL-\rhoG)^{\frac{d-1}{d}}}{2\mykappa \tauW} \left(\rho -
  \rho_0\right)^{\frac{d+1}{d}}~V^{\frac{1}{d}}.
  \label{eqCondensationFixTDelta}
\end{equation}
At fixed temperature $\rhoL,\rho_0,\mykappa,\tau_W$ are constants and $\Delta$
may be interpreted as an unusual density. In principle, all constants may be
estimated and for the present case, equivalent to the 2D Ising model, the
parameters are even known exactly or with very high precision.

This leading-order formulation allows one to obtain the fraction of particles inside
the largest droplet $\lambda$ as a function of $\Delta$ in the limit of large
systems, by minimizing Eq.~(\ref{eqCondensationFixTFreeEnergy}) with respect to
$\lambda$.
It turns out (for details see Refs.~\cite{biskup2002, biskup2003}) that there
exists a constant $\DeltaC$ below which no droplet forms ($\lambda=0$) and above
which a single macroscopic droplet exists with non-trivial $\lambda>\lambdaC$:
\begin{equation}
  \DeltaC  = \frac{1}{d}\left(\frac{d+1}{2}\right)^{\frac{d+1}{d}}
  \overset{\mbox{\normalfont\tiny\sffamily 2D}}{=}0.9186...
  \qquad
  \text{and}
  \qquad
  \lambdaC = \frac{2}{d+1}
  \overset{\mbox{\normalfont\tiny\sffamily 2D}}{=} 2/3.
  \label{eqCondensationFixTDeltaC}
\end{equation}
The result $\lambda(\Delta)$ describes the expectation value of the equilibrium
droplet size in the limit of large systems without any free parameter. As
mentioned before, this already includes the leading-order finite-size
corrections for idealized assumptions. In fact,
Eq.~(\ref{eqCondensationFixTDelta}) may be rewritten at each finite-size
transition density $\rhoC$, where $\Delta(\rhoC)=\DeltaC$. For a lattice gas
model with particle-hole symmetry, $\rhoL=1-\rhoG$, this yields to leading order
\begin{equation}
  \rhoC = \rho_0 +
  \left(\frac{2\mykappa\tauW\DeltaC}{(1-2\rhoG)^{\frac{d-1}{d}}}\right)^{\frac{d}{d+1}}~V^{-\frac{1}{d+1}},
  \qquad\text{or}\qquad
  m_{\rm c} = m_0 -
  2m_0\left(\frac{\chi\tauWI\DeltaC}{2m_0^2}\right)^{\frac{d}{d+1}}~V^{-\frac{1}{d+1}},
  \label{eqCondensationFixTCorrections}
\end{equation}
where the second formulation is in terms of the Ising model at fixed
magnetization. This is the notation of Biskup et al.~\cite{biskup2002,
biskup2003}, but is in quantitative agreement with the (independent) result of
Neuhaus and Hager~\cite{neuhaus2003}.  For the 2D Ising model, they find the
same leading scaling behavior \mbox{$\Delta m(L)=A_{\rm cond}L^{-2/3}$}, with
the amplitude \mbox{$A_{\rm cond}=0.23697...$} for \mbox{$\betaI=0.7$} and
Eq.~\eqref{eqCondensationFixTCorrections} yields $A=0.236965...$ for the
corresponding constants.\footnote{%
  For $\betaI=0.7$ we obtain $m_0\simeq0.99016$, $\chi\simeq0.019310$ and
  $\tauWI\simeq4.5758$.
} 

\begin{figure}
  \centering
  \includegraphics{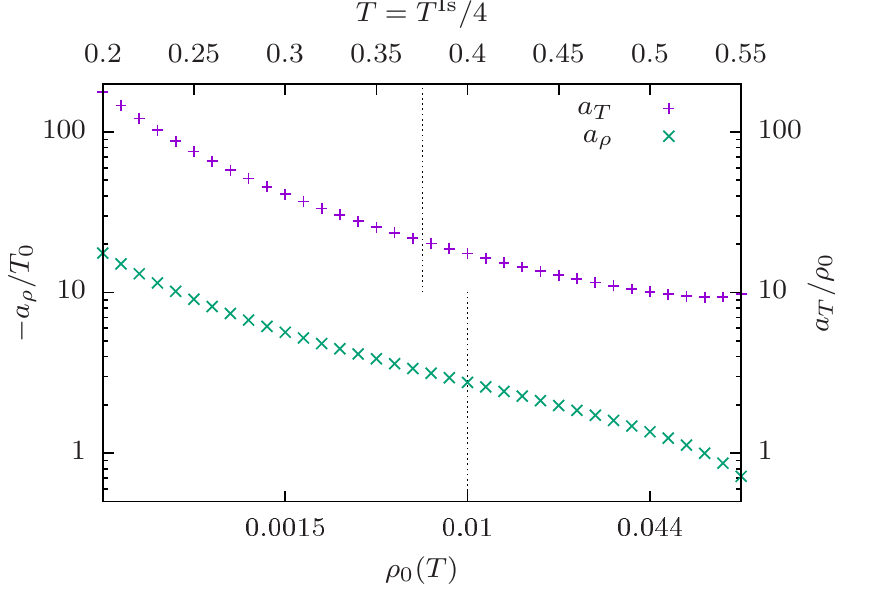}
  \caption{%
    Leading-order normalized finite-size correction amplitude at fixed
    temperature $T$ and fixed density $\rho$. The densities $\rho_0(T)$ on the
    lower x-axis are calculated from the Onsager-Yang solution Eq.~\eqref{eqM}. The
    data points are obtained by numerically evaluating
    Eq.~\eqref{eqFitRelationFixT} and Eq.~\eqref{eqFitRelationFixRho}. Dashed
    lines indicate the parameters considered below.
    \label{figAnalytic}
  }
\end{figure}

%%%%%%%%%%%%%%%%%%%%%%%%%%%%%%%%%%%%%%%%%%%%%%%%%%%%%%%%%%%%%%%%%%%%%%%%%%5
%%%% scope of the paper
In the following, we will numerically review the leading-order finite-size
scaling predictions. We consider directly the scaling of the factual transition
density (or magnetization) at fixed temperature~\cite{nussbaumer2006,
nussbaumer2008} and compare to the scaling of the transition temperature at
fixed density~\cite{zierenberg2015}. Albeit the possibility of logarithmic
corrections, we consider empirical higher-order corrections as powers of the
leading term if necessary.  Our fit ansatz for the leading-order correction is 
\begin{align}
  \rho_c &= \rho_0   + a_{T}    V^{-1/3} + \mathcal{O}\left(V^{-2/3}\right)
         &&\mbox{at fixed $T$}    \label{eqFitFixT},\\
  T_c    &= T_0 + a_{\rho} V^{-1/3} + \mathcal{O}\left(V^{-2/3}\right)
         &&\mbox{at fixed $\rho$} \label{eqFitFixRho}.
\end{align}
%
%a_T
By comparing to Eq.~(\ref{eqCondensationFixTCorrections}), $a_{T}$ can be
related to $\DeltaC$.
%a_rho
Similarly, a relation for $a_{\rho}$ follows from a Taylor expansion around
$T_0$ of a reformulated Eq.~(\ref{eqCondensationFixTDelta}), namely
$\Delta^{2/3} V^{-1/3} = f(\rho,T)$. From the equivalence to the Ising model,
we identify $f(\rho,T)$ with 
$f(m,\TI) = (m_0(\TI)-m)[m_0(\TI)/2]^{1/3}[\chi(\TI)\tauWI(\TI)]^{-2/3}$
(see Ref.~\cite{zierenberg2015} for details), which leads to
\begin{align}
  \DeltaC &= a_T^{3/2} 2m_0^{1/2}/\chi\tauWI
          &&\text{at fixed $T$}  \label{eqFitRelationFixT},\\
  \DeltaC &= \left[a_{\rho}4f'(m,\TI_0)\right]^{3/2}
          &&\text{at fixed $\rho$} \label{eqFitRelationFixRho}.
\end{align}
%
% relative leading-order
This may be evaluated numerically considering Eqs.~\eqref{eqM}--\eqref{eqW} and
\eqref{eqCondensationFixTDeltaC} to yield estimates of the leading-order
correction. Figure~\ref{figAnalytic} shows results for a range of temperatures
and according densities, related by the Onsager-Yang solution Eq.~\eqref{eqM}. The
relative leading-order corrections at fixed temperature are almost an order of
magnitude larger than at fixed density.
% however higher order is relevant
However, in the practical finite-size scaling analysis higher-order corrections
are of more significance, which cannot be estimated by the current theory.

%%%%%%%%%%%%%%%%%%%%%%%%%%%%%%%%%%%%%%%%%%%%%%%%%%%%%%%%%%%%%%%%%%%%%%%%%%%%%%%%%%%%%%%%%%%%
%
% ######
% #     # ######  ####  #    # #      #####  ####
% #     # #      #      #    # #        #   #
% ######  #####   ####  #    # #        #    ####
% #   #   #           # #    # #        #        #
% #    #  #      #    # #    # #        #   #    #
% #     # ######  ####   ####  ######   #    ####
%
%%%%%%%%%%%%%%%%%%%%%%%%%%%%%%%%%%%%%%%%%%%%%%%%%%%%%%%%%%%%%%%%%%%%%%%%%%%%%%%%%%%%%%%%%%%%
\section{Results}

\begin{figure}
  \centering
  \mbox{
    \includegraphics{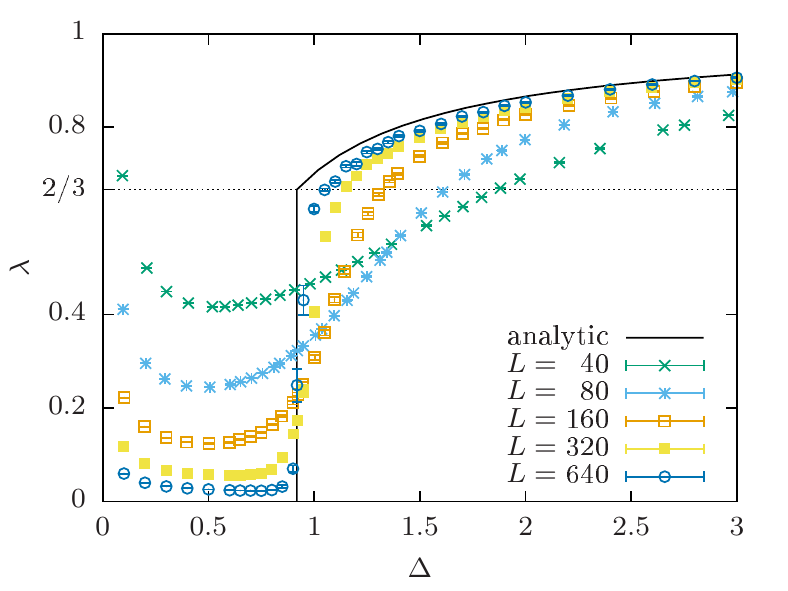}
    \hfill
    \includegraphics{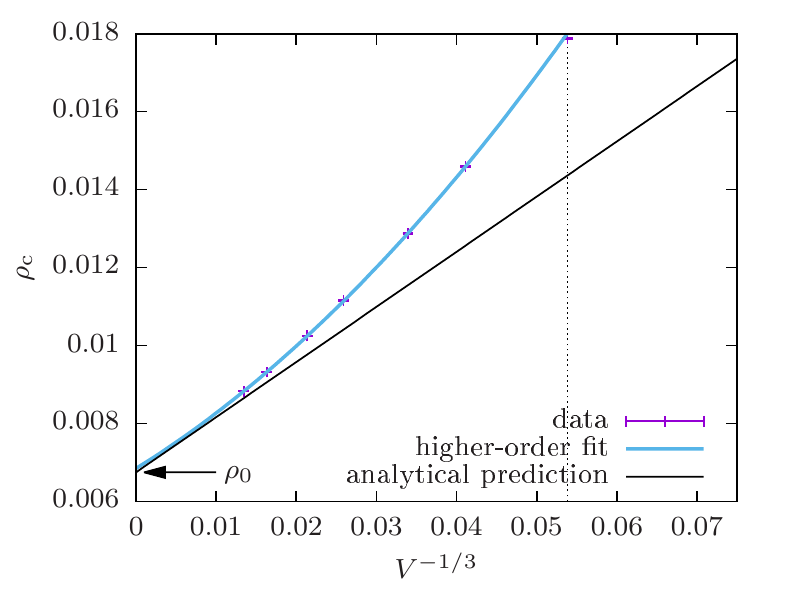}
  }

  \caption{Droplet formation at fixed temperature $\kB
    T=0.375$~\cite{nussbaumer2006,nussbaumer2008}. (left) Fraction of excess
    versus rescaled dimensionless density $\Delta$. (right) Finite-size scaling
    of the transition density $\rhoC$. The black line shows the predicted
    leading-order scaling and the blue line is a higher-order fit to the data.
    $\rho_0$ is the analytically known thermodynamic limit. The dotted vertical
    line indicates the end of the fitting range ($L=80$).
    \label{figFixTFss}
  }
\end{figure}

We will start the discussion with our findings at fixed temperature $\kB
T=0.375$.  Most results are analyses of Monte Carlo time series data obtained by
refined methods analogous to Refs.~\cite{nussbaumer2006,nussbaumer2008}. For
each system size, we set up individual simulations at selected densities around
the expected transition region, see Fig.~\ref{figFixTFss}~(left). For densities
below the finite-size transition point $\Delta<\DeltaC$ the system is in the
gaseous phase and for $\Delta>\DeltaC$ a single macroscopic droplet forms in
accordance with theory.
In fact, the results are obtained in the equivalent formulation of an Ising
model at fixed magnetization, but here discussed in the generic formulation of a
lattice gas. In order to ensure ergodic sampling, we augmented the Kawasaki
Monte Carlo update with a multicanonical scheme to locally sample a flat
histogram including transition states and ensuring a good sampling of the energy
probability distribution up to suppressions of 30 orders of magnitude.
% lambda
The fraction of excess in the largest droplet, Eq.~\eqref{eqLambda}, is obtained
by a two-step process. Firstly, all clusters in the system are determined, where
we define a cluster as alike sites connected via the nearest-neighbor property.
The largest cluster is the background (or gaseous phase), while the second
largest cluster forms the droplet we are interested in.  Then, the volume of the
droplet is identified as all sites confined by the boundary, including the
enclosed holes. The fraction $\lambda$ is then the ratio of the droplet volume
and the expected equilibrium volume of full excess
$V_\delta=(N-\rho_0L^2)/(1-2\rho_0)$. The result of this procedure is shown in
Fig.~\ref{figFixTFss}~(left). We also show the limiting case that can be found
as the solution of Eq.~\eqref{eqCondensationFixTFreeEnergy} as solid black line.
Every data point is the average of $10^6$ Monte Carlo sweeps ($L^2$ updates). At
the infinite-size transition density, the analytical prediction is
$\lambdaC=2/3$, see Eq.~\eqref{eqCondensationFixTDeltaC}. Hence, we estimate the
finite-size transition point as the density $\rhoC$ for which
$\lambda(\rhoC,L)=2/3$.  Technically, we do a linear interpolation of the data
points in Fig.~\ref{figFixTFss}~(left) and search for the intersection.

The resulting scaling of the transition density is shown in
Fig.~\ref{figFixTFss}~(right) in dependence on $V^{-1/3}$ as expected from
Eq.~\eqref{eqFitFixT}. The thermodynamic limit is given by the Onsager solution,
$\rho_0 \simeq 0.00675$. From the analytical prediction in
Eq.~\eqref{eqFitRelationFixT}, we expect for the leading-order correction at
fixed temperature $\kB T=0.375$ that $\DeltaC\simeq0.919\approx(6.6833
~a_T)^{3/2}$ and thus $a_T\approx0.141$,%
\footnote{%
  For $\kB
  T=\kB\TI/4=0.375$ we obtain $m_0\simeq0.9865$ or $\rho_0\simeq0.00675$,
  $\chi=\mykappa\simeq0.02708$, and $\tauWI\simeq4.2454$ or $\tauW\simeq1.0613$.
}
shown in the figure as a black line.
%fit
The data does not allow for a qualitatively satisfying leading-order fit, but
the largest system size is already close to the predicted finite-size deviation.
Considering the next higher empirical correction, i.e., $\rho_c=\rho_0 +
a_TV^{-1/3}+b_TV^{-2/3}$, allows for a decent fit which yields for $L>80$ the
result $\rho_0=0.00676(5)$ and $a_T=0.135(3)$ with goodness-of-fit parameter
$Q\approx0.02$. The estimated limit is in good agreement with the analytical
prediction $\rho_0$ marked by the arrow. Using Eq.~(\ref{eqFitRelationFixRho})
with error propagation yields the estimate $\Delta_c=0.86(3)$, which is below
but consistent with the theoretical prediction. We notice that the choice of the
intersection in $[0,2/3]$ strongly influences the size of the higher-order
corrections.

\begin{figure}
  \centering
  \mbox{
    \includegraphics{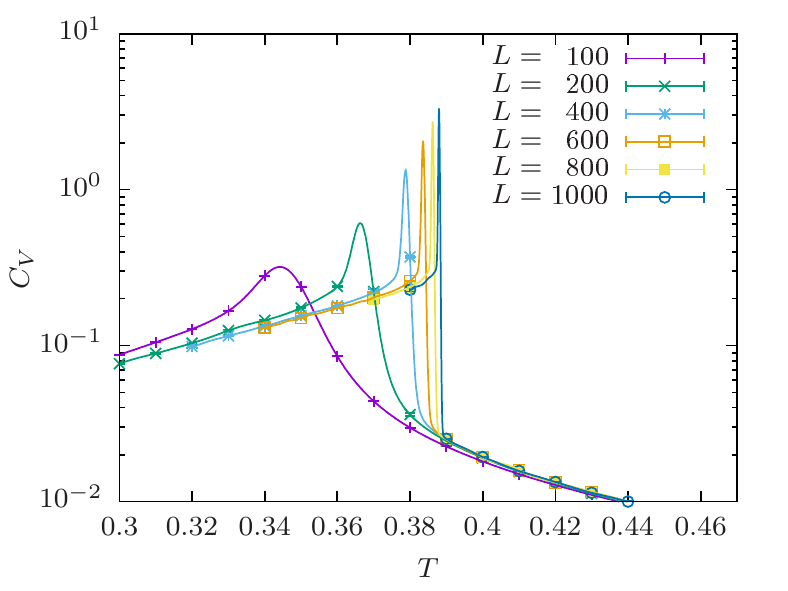}
    \hfill
    \includegraphics{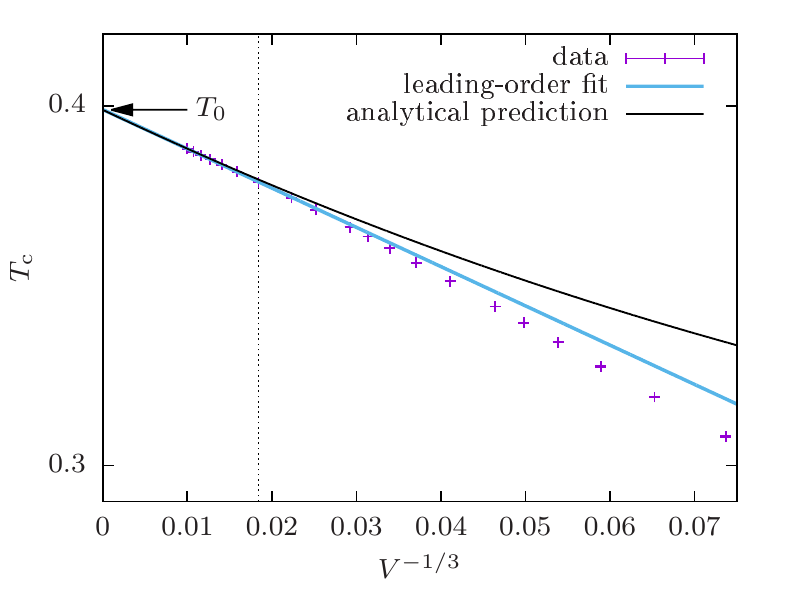}
  }
  \caption{%
    Droplet formation at fixed density $\rho=10^{-2}$~\cite{zierenberg2015}.
    (left) Specific heat with exemplary data points that indicate the size of
    the error.
    (right) Finite-size scaling of the transition temperature. The black line
    shows the numerical evaluation of Eq.~(\ref{eqCondensationFixTDelta}) at
    fixed $\rho$ and the blue line is a leading-order fit. $T_0$ is the
    analytically known thermodynamic limit. The dotted vertical line indicates
    the end of the fitting range ($L=400$).
    \label{figFixRhoFss}
  }
\end{figure}
We next turn to the orthogonal setup of a fixed density \mbox{$\rho=10^{-2}$}
with varying temperature~\cite{zierenberg2015}. Here, the multicanonical method
is more straightforward and for each system size we may perform a single (yet
parallel) simulation with up to 128 cores and $1.28\times10^{6}$ measurements in
the final production run. This yields a full temperature range of expectation
values and the transition temperature may be determined precisely as the peak of
the specific heat $C_V=\kB\beta^2\left(\langle E^2\rangle -\langle
E\rangle^2\right)/N$, see Fig.~\ref{figFixRhoFss}~(left). For details see
Ref.~\cite{zierenberg2015}, from which we recapture the data in order to compare
to the leading-order results at fixed temperature. For $\rho=10^{-2}$, the
numerical evaluation of the Onsager solution Eq.~\eqref{eqM} yields
$T_0\simeq0.39882$, which is the expected thermodynamic limit. From the
analytical prediction Eq.~\eqref{eqFitRelationFixRho}, we expect for the
leading-order correction that
$\DeltaC\simeq0.919\approx(-0.85177~a_{\rho})^{3/2}$ and thus
$a_{\rho}\approx-1.109$.

% figure and leading-solution
Figure~\ref{figFixRhoFss}~(right) shows the finite-size transition temperature
as a function of the expected leading-order scaling correction. In two
dimensions, we can numerically evaluate Eq.~(\ref{eqCondensationFixTDelta}) at
fixed density for various system sizes. This gives the (analytical) leading-order
finite-size estimate of the transition temperature, shown as a black line. We
can see that this predicted leading-order behavior is only approximately reached
for very large system sizes of about $L\simeq500$. For small system sizes, the
leading-order solution strongly deviates, which may be expected from the
simplified assumptions.
%
% fit
% L=200->N=400
% L=500->N=2500
A leading-order fit for $L\ge400$ yields $T_0=0.39891(5)$ and
$a_{\rho}=-1.091(4)$ with $Q\approx0.25$. The limit is in good agreement with
the analytical value. Using Eq.~(\ref{eqFitRelationFixRho}) with error
propagation yields the estimate $\DeltaC=0.896(5)$. This is close to, but
differs slightly from, the predicted value. As for the case of fixed
temperature, this is below the theoretical prediction. The fit error is too
small in order to be fully consistent, which may be accounted to missing
higher-order corrections. We also tried the empirical higher-order corrections,
however, it did not improve our estimates. This may be explained by additional
corrections, e.g., logarithmic ones, which were not considered. Still, the
qualitatively good leading order fit shows that in this setup we are close to
the asymptotic scaling regime.

%%%%%%%%%%%%%%%%%%%%%%%%%%%%%%%%%%%%%%%%%%%%%%%%%%%%%%%%%%%%%%%%%%%%%%%%%%%%%%%%%%%%%%%%%%%%
%
%  #####
% #     #  ####  #    #  ####  #      #    #  ####  #  ####  #    #
% #       #    # ##   # #    # #      #    # #      # #    # ##   #
% #       #    # # #  # #      #      #    #  ####  # #    # # #  #
% #       #    # #  # # #      #      #    #      # # #    # #  # #
% #     # #    # #   ## #    # #      #    # #    # # #    # #   ##
%  #####   ####  #    #  ####  ######  ####   ####  #  ####  #    #
%
%%%%%%%%%%%%%%%%%%%%%%%%%%%%%%%%%%%%%%%%%%%%%%%%%%%%%%%%%%%%%%%%%%%%%%%%%%%%%%%%%%%%%%%%%%%%
\section{Conclusions} 
We numerically investigated the leading-order finite-size scaling corrections
of the two-dimensional droplet condensation-evaporation transition at fixed
temperature and density, respectively. In both cases we could recover the
analytically known thermodynamic limits, in part by considering empirical
higher-order corrections.  The leading-order corrections were found to be
slightly smaller than but consistent with the analytical
predictions~\cite{biskup2002,biskup2003}. This shows that the established
theory is able to qualitatively describe the finite-size deviations; in case of
the two-dimensional lattice gas already for system sizes $L\simeq500$. 

%%%%%%%%%%%%%%%%%%%%%%%%%%%%%%%%%%%%%%%%%%%%%%%%%%%%%%%%%%%%%%%%%%%%%%%%%%%%%%%%%%%%%%%%%%%%
%%%%%%%%%%%%%%%%%%%%%%%%%%%%%%%%%%%%%%%%%%%%%%%%%%%%%%%%%%%%%%%%%%%%%%%%%%%%%%%%%%%%%%%%%%%%
\section*{Acknowledgments}
The project was funded by the European Union and the Free State of Saxony.
This work has been partially supported by
  the DFG (Grant No.\ JA 483/31-1),
  the Leipzig Graduate School ``BuildMoNa'',
  and the Deutsch-Franz\"osische Hochschule DFH-UFA (Grant No.\ CDFA-02-07).
The authors gratefully acknowledge the computing time provided by the John von
Neumann Institute for Computing (NIC) on the supercomputer JUROPA at J\"ulich
Supercomputing Centre (JSC).

\bibliographystyle{iopart-num}
\section*{References}
\bibliography{references}

\providecommand{\newblock}{}
\begin{thebibliography}{10}
\expandafter\ifx\csname url\endcsname\relax
  \def\url#1{{\tt #1}}\fi
\expandafter\ifx\csname urlprefix\endcsname\relax\def\urlprefix{URL }\fi
\providecommand{\eprint}[2][]{\url{#2}}

\bibitem{binder1980}
Binder K and Kalos M~H 1980 {\em J. Stat. Phys.\/} {\bf 22} 363

\bibitem{neuhaus2003}
Neuhaus T and Hager J 2003 {\em J. Stat. Phys.\/} {\bf 113} 47

\bibitem{biskup2002}
Biskup M, Chayes L and Koteck{\'y} R 2002 {\em Europhys. Lett.\/} {\bf 60} 32

\bibitem{biskup2003}
Biskup M, Chayes L and Koteck{\'y} R 2003 {\em J. Stat. Phys.\/} {\bf 116} 175

\bibitem{binder2003}
Binder K 2003 {\em Physica A\/} {\bf 319} 99

\bibitem{nussbaumer2006}
Nu{\ss}baumer A, Bittner E, Neuhaus T and Janke W 2006 {\em Europhys. Lett.\/}
  {\bf 75} 716

\bibitem{nussbaumer2008}
Nu{\ss}baumer A, Bittner E and Janke W 2008 {\em Phys. Rev. E\/} {\bf 77}
  041109

\bibitem{zierenberg2014jpcs}
Zierenberg J, Wiedenmann M and Janke W 2014 {\em J. Phys: Conf. Ser.\/} {\bf
  510} 012017

\bibitem{macdowell2004}
MacDowell L~G, Virnau P, M{\"u}ller M and Binder K 2004 {\em J. Chem. Phys.\/}
  {\bf 120} 5293

\bibitem{macdowell2006}
MacDowell L~G, Shen V~K and Errington J~R 2006 {\em J. Chem. Phys.\/} {\bf 125}
  034705

\bibitem{schrader2009}
Schrader M, Virnau P and Binder K 2009 {\em Phys. Rev. E\/} {\bf 79} 061104

\bibitem{martinos2007}
Martinos S, Malakis A and Hadjiagapiou I 2007 {\em Physica A\/} {\bf 384} 368

\bibitem{zierenberg2015}
Zierenberg J and Janke W 2015 {\em Phys. Rev. E\/} {\bf 92} 012134

\bibitem{lee1952}
Lee T~D and Yang C~N 1952 {\em Phys. Rev.\/} {\bf 87} 410

\bibitem{onsager1949}
Onsager L 1949 {\em Nuovo Cimento (Suppl.)\/} {\bf 6} 261

\bibitem{yang1952}
Yang C~N 1952 {\em Phys. Rev.\/} {\bf 85} 808

\bibitem{boukraa2008jphysA}
Boukraa S, Guttmann A~J, Hassani S, Jensen I, Maillard J~M, Nickel B and Zenine
  N 2008 {\em J. Phys. A: Math. Theor.\/} {\bf 41} 455202

\bibitem{leung1990}
Leung K and Zia R~K~P 1990 {\em J. Phys. A: Math. Gen.\/} {\bf 23} 4593

\bibitem{berg1991}
Berg B~A and Neuhaus T 1991 {\em Phys. Lett. B\/} {\bf 267} 249

\bibitem{berg1992}
Berg B~A and Neuhaus T 1992 {\em Phys. Rev. Lett.\/} {\bf 68} 9

\bibitem{janke1992}
Janke W 1992 {\em Int. J. Mod. Phys. C\/} {\bf 3} 1137

\bibitem{janke1998}
Janke W 1998 {\em Physica A\/} {\bf 254} 164

\bibitem{zierenberg2013cpc}
Zierenberg J, Marenz M and Janke W 2013 {\em Comput. Phys. Commun.\/} {\bf 184}
  1155

\bibitem{zierenberg2014pp}
Zierenberg J, Marenz M and Janke W 2014 {\em Physics Procedia\/} {\bf 53} 55

\bibitem{efron1982}
Efron B 1982 {\em The Jackknife, the Bootstrap and Other Resampling Plans\/}
  (Society for Industrial and Applied Mathematics)

\end{thebibliography}

\end{document}